\documentclass[aps,prc,reprint,superscriptaddress]{revtex4-2}
\usepackage{amsmath}
\usepackage{graphicx}
\usepackage{microtype}
\usepackage{placeins}
\begin{document}
\title{UrQMD-Agent: An LLM-Assisted Framework for Microscopic Tracing of Proton Azimuthal Correlations}
\author{Fu-Peng Li}
\email[]{fpli@fudan.edu.cn}
\affiliation{Key Laboratory of Nuclear Physics and Ion-beam Application (MOE) \& Institute of Modern Physics, Fudan University, Shanghai 200433, China}
\affiliation{Shanghai Research Center for Theoretical Nuclear Physics,
NSFC and Fudan University, Shanghai 200438, China}
\author{Yige Huang}
\email[]{huangyige@impcas.ac.cn}
\affiliation{State Key Laboratory of Heavy Ion Science and Technology, Institute of Modern Physics, Chinese Academy of Sciences, Lanzhou 730000, China}

\begin{abstract}
We present an LLM-assisted framework for microscopic transport studies that connects a general-purpose large language model (LLM) to the UrQMD transport model through a model-specific Skill and a deterministic scientific tool layer. The LLM supports literature-guided scientific planning and interpretation, while numerical calculations, event-history reconstruction, and statistical analyses are performed by deterministic tools. As a demonstration, the framework is applied to proton azimuthal-partition fluctuations in Au+Au collisions at $\sqrt{s_{NN}}=3.0$--39~GeV. In the 0--5\% centrality class, the fluctuations are statistically resolved below zero from 3 to 27~GeV and consistent with zero at 39~GeV. An exact pair-kernel decomposition shows that the negative signal is carried mainly by proton pairs containing at least one decay-fed proton. A random-label null test further distinguishes the observed pair contributions from expectations based on pair abundance alone. Reaction-history tracing shows that, for pairs of two decay-fed protons, the negative pair-kernel contribution is already present in the parent momenta evaluated at their respective formation vertices. The modular architecture is designed to accommodate different general-purpose LLMs and can be extended to other microscopic transport models through dedicated Skills and tool interfaces. This provides a pathway toward systematic microscopic analyses and comparisons across transport models within a common LLM-assisted framework.
\end{abstract}

\maketitle

\section{Introduction}
\label{sec:intro}

Large language models (LLMs) are increasingly employed as scientific agents that integrate natural-language reasoning with literature retrieval, code execution, and domain-specific computational tools~\cite{team2025kimi,liu2024deepseek,ling2026agent,wang2024survey,roumeliotis2023chatgpt,chang2024survey,Boiko2023,Bran2024,Lu2024}. Recent applications in physics include automated Monte Carlo workflows in nuclear engineering, quantum many-body simulations and tensor-network algorithm development, collider phenomenology and analysis reproduction, experimental high-energy physics analyses, and relativistic heavy-ion hydrodynamics~\cite{He:2026jcq,Esmail:2026jpb,miao2025physmaster,AutoFLUKA2025,QuantumAgent2026,PhysVEC2026,QuantumWorkflow2026,ColliderAgent2026,ColliderBench2026,LLM4HEP2025,JFC2026,DrSai2026,CLViscAgent2026,Menzo:2026xat,Desai:2026nmx}. These studies illustrate the use of LLM agents in computational workflows involving model configuration, simulation, data processing, and interpretation.

Relativistic heavy-ion collisions provide an important means of investigating the properties and dynamical evolution of strongly interacting matter~\cite{BraunMunzinger2016,HeinzSnellings2013,Gross:2022hyw,Sorensen:2023zkk,Du:2024wjm,Busza:2018rrf,Elfner:2022iae,STAR:2020tga,STAR:2021iop,Chen:2024aom,Pang:2012he,Pang:2018zzo}. Microscopic transport models describe the collision dynamics at the particle level by following particle propagation, scattering, production, and decay processes on an event-by-event basis~\cite{Bass1998,Bleicher:1999xi,Cassing:2009vt,Lin:2004en,SMASH:2016zqf,TMEP:2017mex}. Among these models, the Ultra-relativistic Quantum Molecular Dynamics (UrQMD) model has been widely used to study hadronic observables over a broad range of collision energies~\cite{Cozma:2013sja,Petersen:2006vm,Wang:2013wca,Wang:2018hsw,Zhang:2025ale,OmanaKuttan:2022aml,Li:2021plq,Li:2022wvu,Li:2022iil,OmanaKuttan:2022the}. When microscopic interactions and decays are recorded, UrQMD also allows final-state observables to be examined in relation to the underlying reaction history. Such analyses require detailed knowledge of the model implementation, reconstruction of particle histories, and dedicated statistical tools.

In this work, we construct an LLM-assisted environment by coupling a general-purpose LLM with a local UrQMD installation through the model-specific UrQMD Skill and a deterministic scientific tool layer. Guided by relevant literature and model documentation, the LLM plans the simulation and analysis workflow, while deterministic tools prepare inputs, execute UrQMD calculations, compute observables, and reconstruct microscopic event histories. Changes to the physics configuration and new production runs require user confirmation. The LLM interprets the validated results, whereas numerical evidence is obtained through deterministic operations on recorded UrQMD output. This framework enables final-state observables to be connected to microscopic reaction histories rather than treating UrQMD solely as an event generator.

We demonstrate the framework by studying proton azimuthal-partition fluctuations in Au+Au collisions. This observable measures the deviation of proton multiplicity fluctuations from an independent-binomial baseline within randomly oriented azimuthal windows, and negative values have been reported in transport calculations~\cite{Neff2024}. Using UrQMD simulations at $\sqrt{s_{NN}}=3.0$, 7.7, 11.5, 19.6, 27, and 39~GeV, we first examine the dependence of the observable on beam energy, centrality, proton multiplicity, and partition width using high-statistics samples. We then introduce an exact eventwise proton-pair representation to identify the pair populations contributing to the negative signal. For pairs of two decay-fed protons (DD), we further use the recorded reaction history to evaluate the pair correlation with the momenta of their parent resonances immediately before decay and at their respective formation vertices.

The remainder of this paper is organized as follows. Section~\ref{sec:method} describes the UrQMD agent framework, simulation setup, observable, and microscopic history analysis. Section~\ref{sec:results} presents the high-statistics benchmark tests and the proton-pair-level reaction-history results. Section~\ref{sec:summary} summarizes the main conclusions and outlines future work.

\section{Method}
\label{sec:method}

\subsection{UrQMD Skill and agentic workflow}
\label{sec:framework}

\begin{figure*}[htbp]
\centering
\includegraphics[width=\textwidth,keepaspectratio]{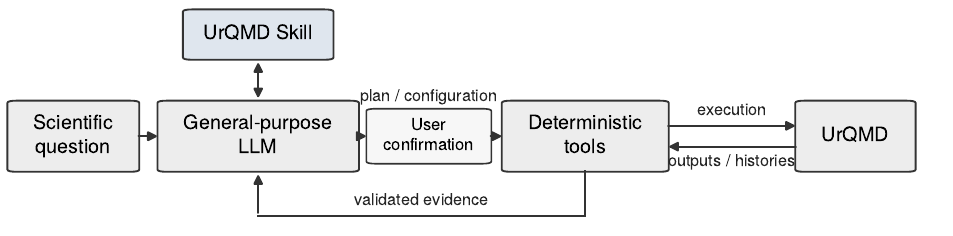}
\caption{\label{fig:framework}The architecture of the LLM+UrQMD analysis
framework. A general-purpose LLM formulates the analysis with
model-specific guidance from the UrQMD Skill. Execution-changing
actions pass through user confirmation before deterministic tools
configure and run UrQMD, process the resulting event records, and
return validated evidence to the LLM. All numerical results are
produced by the deterministic tool layer.}
\end{figure*}

Figure~\ref{fig:framework} illustrates the LLM-assisted analysis framework employed in this work, which connects a general-purpose LLM to a local UrQMD installation through a model-specific UrQMD Skill and a deterministic scientific tool layer. Given a scientific question, the LLM consults relevant literature and model documentation and uses the UrQMD Skill to propose appropriate model settings and analysis procedures. The Skill provides model-specific knowledge and guidance, including simulation conventions, configuration options, analysis procedures, and validation rules grounded in the UrQMD source code and documentation. After the proposed physics configuration and computational plan are confirmed by the user, the LLM invokes the deterministic tools to prepare input files, execute UrQMD simulations, and process the resulting event records. These tools support the calculation of final-state observables, pair-level correlation decomposition, reconstruction of particle genealogies and microscopic reaction histories, statistical analyses and consistency checks, and controlled transport interventions. All numerical computations are performed by these tools rather than by the LLM itself, supporting the reproducibility and traceability of the analysis. The validated results are then returned to the LLM for scientific interpretation and the formulation of subsequent analyses. In the present study, this framework is first used to evaluate proton azimuthal-partition fluctuations with high-statistics UrQMD event samples and then to trace the observed negative correlation through the recorded microscopic reaction history.

\subsection{UrQMD setup and event samples}
\label{sec:samples}

We simulated Au+Au collisions at $\sqrt{s_{NN}}=3.0$, 7.7, 11.5, 19.6, 27, and 39~GeV using the UrQMD transport model in cascade mode. For the azimuthal-partition analysis, we generated 200{,}000 events at each energy. In addition, a separate set of 80{,}000 events per energy was generated with the full microscopic interaction history preserved, for pair decomposition and reaction history tracking. Protons were selected in the midrapidity region with $|y|<0.5$ and $0.4<p_T<2$~GeV/$c$. Event activity is characterized by RefMult, defined as the number of final-state charged hadrons within $|\eta|<0.5$. Unless otherwise stated, only interacting events—those in which at least one binary nucleon--nucleon collision occurs—are included in the analysis.

The 200,000-event benchmark analysis uses the original RefMult definition, while the microscopic analysis uses a proton-excluded reference multiplicity, ${\rm RefMult}3$, defined by excluding protons and antiprotons from the charged-hadron count within $|\eta|<0.5$. A cross-check within the history-rich sample shows that the total fluctuation in the 0–5\% centrality class is statistically consistent between the two definitions.

\subsection{Azimuthal-partition observable and statistical analysis}
\label{sec:observable}

The normalized azimuthal-partition fluctuation is defined as
\begin{equation}
\Delta\sigma^2(N,w)=
\frac{\sigma^2(N_w)-N P_0(1-P_0)}{N(N-1)},
\qquad
P_0=\frac{w}{2\pi},
\label{eq:dsigma}
\end{equation}
where $N$ is the accepted-proton multiplicity, $w$ is the azimuthal window width, and $N_w$ is the number of accepted protons within a randomly oriented window of width $w$. Here, $\sigma^2(N_w)$ denotes the variance of $N_w$ at fixed $N$, and $P_0$ is the probability for a proton to fall within the window under a uniform azimuthal distribution. Positive values of $\Delta\sigma^2$ indicate fluctuations above the independent-binomial baseline, while negative values indicate suppressed fluctuations.

For each event, 72 random azimuthal orientations are sampled. The
physical event is taken as the statistical unit, with all orientations
from the same event kept together in the bootstrap resampling.
Statistical uncertainties are obtained with a Poisson event-block
bootstrap and are quoted as one bootstrap standard error. Statistical
resolution is assessed from the corresponding 95\% bootstrap interval.
For the microscopic history-rich analysis, an ordinary event-block
bootstrap with 2000 replicas is used, and paired quantities evaluated
at different genealogical levels share the same resampled event
indices, so that stage-to-stage contrasts retain their event-level
correlations.

Results are presented both at fixed $N$ and after averaging over the
supported $N$ bins. The fixed-$N$ analysis is performed at
$w=120^\circ$, while the event-activity dependence is shown as a
function of RefMult. The width dependence is evaluated at
$w=60^\circ$, $72^\circ$, $90^\circ$, $120^\circ$, $180^\circ$,
$240^\circ$, $270^\circ$, $288^\circ$, and $300^\circ$, using a common
fixed-$N$ weighting defined at $w=120^\circ$. Only bins with at least
20 contributing events are included in the averaged results. The width
dependence is summarized with the empirical form~\cite{Neff2024}
\begin{equation}
\langle\Delta\sigma^2\rangle(w)=
b\left[
1-\left(\frac{w-180^\circ}{z}\right)^2
\right],
\label{eq:bz}
\end{equation}
where $b$ characterizes the amplitude and $z$ the curvature of the
width dependence.

For the microscopic analysis, an exact eventwise representation of the
observable is used. Consider an event with $N$ accepted protons at
azimuthal angles $\phi_1,\ldots,\phi_N$, while the partition orientation
$\psi$ is uniformly sampled over $[0,2\pi)$. Let
$X_i(\psi)\in\{0,1\}$ indicate whether proton $i$ lies inside the
azimuthal window $W_\psi$. Then
$N_w(\psi)=\sum_i X_i(\psi)$,
$\langle X_i\rangle_\psi=P_0$, and
${\rm Var}_\psi(X_i)=P_0(1-P_0)$.
For a proton pair $(i,j)$, the minimal circular separation is
\[
d_{ij}=\min\left(|\phi_i-\phi_j|,
2\pi-|\phi_i-\phi_j|\right).
\]
The probability that both protons lie inside the same window is purely
geometrical,
\begin{equation}
q(w,d)=
\begin{cases}
\max(w-d,0)/(2\pi), & w\le\pi,\\[2mm]
1-\left[2\ell-\max(\ell-d,0)\right]/(2\pi), & w>\pi,
\end{cases}
\label{eq:kernel}
\end{equation}
where $\ell=2\pi-w$. The corresponding pair covariance under random
partition orientations is
\begin{equation}
K(w,d)\equiv {\rm Cov}_\psi(X_i,X_j)
=q(w,d)-P_0^2.
\label{eq:kercov}
\end{equation}
Here $K$ is a geometrical covariance associated with the random
partition orientation, rather than a dynamical interaction between the
two protons. Expanding ${\rm Var}_\psi(N_w)$ gives the exact eventwise
identity
\begin{equation}
\Delta\sigma^2_{\rm event}(w)
=
\frac{2}{N(N-1)}
\sum_{i<j}K(w,d_{ij}).
\label{eq:pairexact}
\end{equation}
Thus, $\Delta\sigma^2$ can be written exactly as the mean pair-kernel contribution over all unordered proton pairs. At fixed $N$, the ensemble average of Eq.~(\ref{eq:pairexact}) reproduces Eq.~(\ref{eq:dsigma}). We verified that the 72-orientation estimator agrees with the analytic pair-kernel evaluation within statistical uncertainties. Events with $N<2$ are excluded from pair-level quantities because the pair normalization is undefined.

Because Eq.~(\ref{eq:pairexact}) is additive in proton pairs, the full
pair sample can be divided into disjoint classes without changing the
normalization. An accepted proton is classified as decay-fed (D) if it
originates from a recorded decay vertex and as non-decay-fed (X)
otherwise. A DD pair contains two decay-fed protons, a DX pair contains
one decay-fed and one non-decay-fed proton, and an XX pair contains two
non-decay-fed protons. Their contributions satisfy
\begin{equation}
\Delta\sigma^2_{\rm total}(w)
=
\Delta\sigma^2_{\rm DD}(w)
+
\Delta\sigma^2_{\rm DX}(w)
+
\Delta\sigma^2_{\rm XX}(w).
\label{eq:additive}
\end{equation}
The decomposition is evaluated at $w=120^\circ$. All DD pairs in the
history-rich samples are associated with two distinct parent resonances.
No same-parent or unresolved-parent DD pairs are present.

An additive class contribution reflects both the abundance of pairs in
that class and their mean correlation strength. For an event with $N$
accepted protons and $M=N(N-1)/2$ unordered pairs, we define the pair
fraction $F_a=n_a/M$ of class $a\in\{\mathrm{DD},\mathrm{DX},
\mathrm{XX}\}$ containing $n_a$ pairs and its per-pair mean kernel
$\bar K_a=n_a^{-1}\sum_{(i,j)\in a}K(120^\circ,d_{ij})$, so that
$\Delta_a=F_a\bar K_a$ holds eventwise. This separation distinguishes
whether a class contributes through pair abundance or through a more
negative per-pair correlation. Ensemble averages are taken over the
eventwise products; $\langle\Delta_a\rangle=
\langle F_a\bar K_a\rangle$ is in general not equal to
$\langle F_a\rangle\langle\bar K_a\rangle$.

To test whether the decay-fed identity is preferentially associated with the pair-kernel structure beyond pair abundance, we construct a random-label null test. For an event with $N_D$ decay-fed and $N_X=N-N_D$ non-decay-fed accepted protons, the class pair counts are $n_{\rm DD}=N_D(N_D-1)/2$, $n_{\rm DX}=N_DN_X$, and $n_{\rm XX}=N_X(N_X-1)/2$. These counts determine the pair fractions $F_a$ defined above. The null preserves the proton azimuthal configuration and the counts $N_D$ and $N_X$ while randomly reassigning the D/X labels among accepted protons. The exact conditional null expectation and the corresponding residual are defined as
\begin{equation}
\Delta^{\rm null}_{a,e}=F_{a,e}\,\Delta_{{\rm total},e},
\qquad
R_{a,e}\equiv\Delta_{a,e}-\Delta^{\rm null}_{a,e},
\label{eq:nullres}
\end{equation}
The total pair contribution remains unchanged, and $R_{{\rm DD},e}+R_{{\rm DX},e}+R_{{\rm XX},e}=0$ holds exactly for each event. A negative $R_a$ indicates a more negative contribution than expected from pair abundance alone, while a positive $R_a$ indicates a less negative or more positive contribution. Ensemble averages of $R_a$ are evaluated using the same event aggregation and paired event-block bootstrap as the pair decomposition.

The recorded reaction histories allow the same DD pair population to be
followed backward through the hadronic evolution. For each decay-fed
proton, the decay record identifies the parent resonance and its
four-momentum immediately before decay. The resonance is then traced
backward through its unique particle identifier to the interaction in
which it was formed. For the identical DD pair population, the exact
pair kernel is evaluated at three genealogical levels $s$,
\begin{equation}
\Delta_{\rm DD}^{s}(w)
=
\frac{2}{N(N-1)}
\sum_{(i,j)\in{\rm DD}}K(w,d_{ij}^{s}),
\label{eq:ddstages}
\end{equation}
where $s\in\{\mathrm{form},\mathrm{predecay},\mathrm{final}\}$ labels
the three genealogical levels corresponding to the parent momenta
evaluated at their respective formation vertices, the parent momenta
immediately before decay, and the final-state proton momenta,
respectively, and $d_{ij}^{s}$ is the
corresponding azimuthal separation of the pair. The final
accepted-proton multiplicity $N$ is used in the normalization at all
three levels, and the construction is evaluated at $w=120^\circ$. The
two parent resonances of a DD pair need not be formed at the same
physical time; the formation-vertex quantity is therefore a
genealogical construction rather than an equal-time pair observable.

The changes between successive levels are defined as
\begin{equation}
\begin{gathered}
\delta_{\rm DD}^{\rm transport}
=
\Delta_{\rm DD}^{\rm predecay}-\Delta_{\rm DD}^{\rm form},\\[1mm]
\delta_{\rm DD}^{\rm decay}
=
\Delta_{\rm DD}^{\rm final}-\Delta_{\rm DD}^{\rm predecay},
\end{gathered}
\label{eq:ddelta}
\end{equation}
so that
$\Delta_{\rm DD}^{\rm form}+\delta_{\rm DD}^{\rm transport}
+\delta_{\rm DD}^{\rm decay}=\Delta_{\rm DD}^{\rm final}$ holds
exactly. The same event resampling is used at all three levels so that
the correlations between the paired quantities are retained. The
tracing uses the recorded decay ancestry and the survival of the
particle identifier through elastic interactions. No parentage is
assigned across generic multiparticle production vertices.

\begin{figure*}[htbp]
\centering
\includegraphics[width=\textwidth,keepaspectratio]{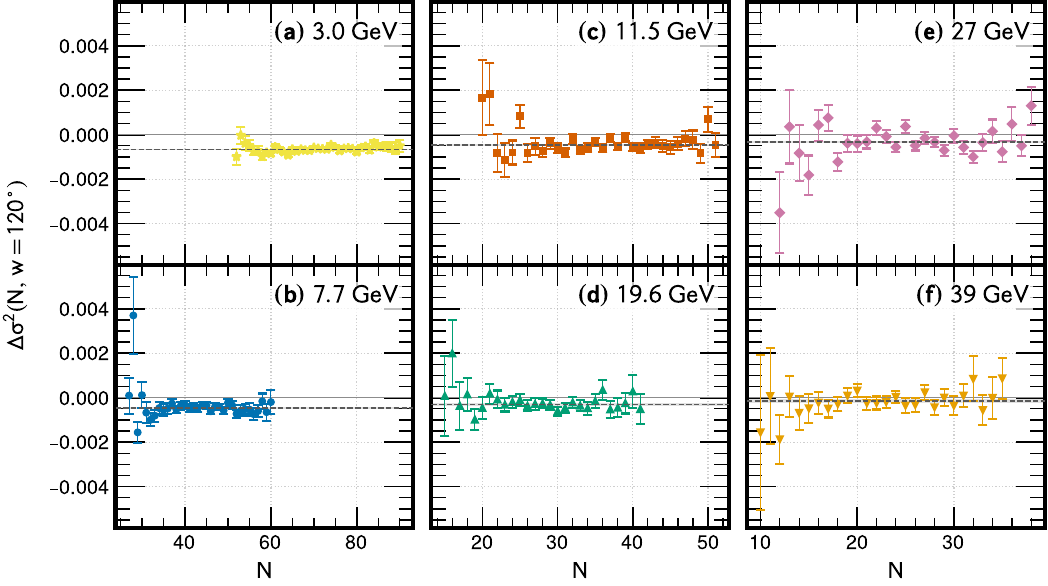}
\caption{\label{fig:fixedN}Proton azimuthal-partition fluctuation
$\Delta\sigma^2(N, w = 120^\circ)$ versus the accepted-proton
multiplicity $N$ in interacting 0--5\% Au+Au collisions at
$\sqrt{s_{NN}} = 3.0$--39~GeV.}
\end{figure*}

\section{Results}
\label{sec:results}
\subsection{Azimuthal-partition from UrQMD}

\begin{figure}[hbtp]
\centering
\includegraphics[width=\columnwidth,keepaspectratio]{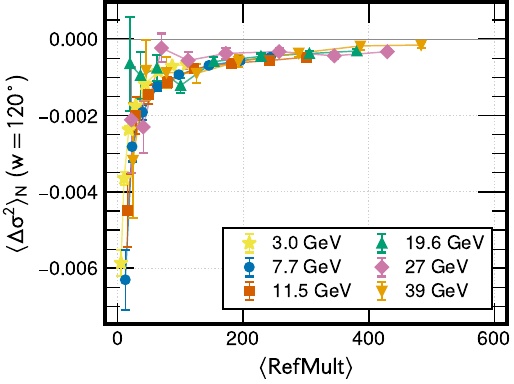}
\caption{\label{fig:refmult}$N$-averaged $\langle\Delta\sigma^2\rangle$
at $w = 120^\circ$ versus the class-mean reference multiplicity RefMult
(charged hadrons at $|\eta| < 0.5$) for centrality classes from 0--5\%
to 60--70\% at six beam energies.}
\end{figure}
\begin{figure*}[htbp]
\centering
\includegraphics[width=\textwidth,keepaspectratio]{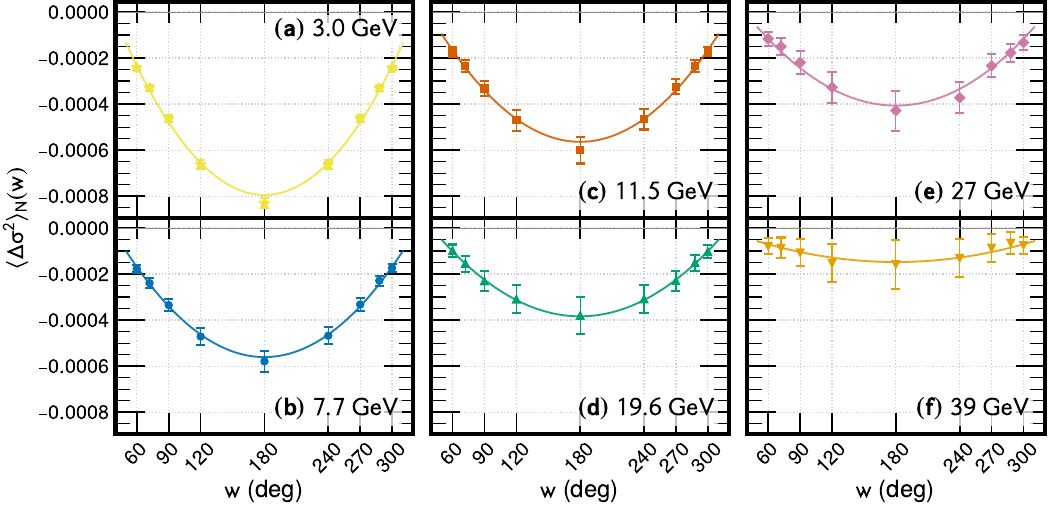}
\caption{\label{fig:width}$\langle\Delta\sigma^2\rangle$ in the 0--5\%
class as a function of the partition width $w$ at six beam energies.}
\end{figure*}

\begin{figure*}[htbp]
\centering
\includegraphics[width=\textwidth,keepaspectratio]{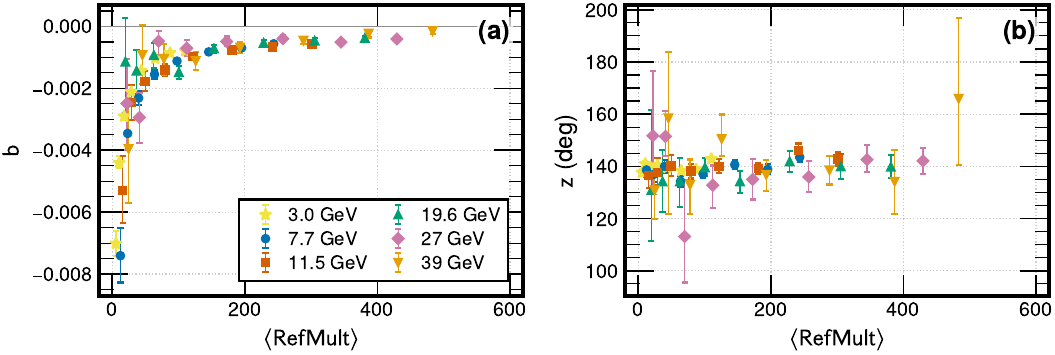}
\caption{\label{fig:bz}Parameters of Eq.~(\ref{eq:bz}) versus the mean RefMult for eight centrality classes at six beam energies: (a) amplitude $b$ and (b) angular-scale parameter $z$. }
\end{figure*}

Figure~\ref{fig:fixedN} shows $\Delta\sigma^2(N,w=120^\circ)$ as a function of the accepted-proton multiplicity $N$ in interacting 0--5\% events at the six beam energies. Negative values are observed over most of the statistically supported fixed-$N$ range. The negative correlation is most pronounced at 3~GeV and becomes substantially weaker toward higher energies, with the results at 39~GeV lying close to zero.

Figure~\ref{fig:refmult} shows $\langle\Delta\sigma^2\rangle$ at
$w=120^\circ$ as a function of the class-mean RefMult for centrality
classes from 0--5\% to 60--70\%. At all six beam energies, the negative
correlation becomes stronger toward smaller RefMult, corresponding to
more peripheral collisions. Its magnitude increases from a few times
$10^{-4}$ in the most central collisions to the $10^{-3}$ level in the
peripheral region. In the 0--5\% class, the signal is statistically
resolved below zero from 3 to 27~GeV, while the 39~GeV result is
consistent with zero. This centrality dependence is considerably
stronger than the variation with fixed proton multiplicity $N$ shown
in Fig.~\ref{fig:fixedN}.

Figure~\ref{fig:width} displays $\langle\Delta\sigma^2\rangle$ as a function of the partition width $w$ for the 0--5\% centrality class. At all six beam energies, the negative correlation is strongest near $w=180^\circ$, with a symmetric dependence about this angle, as expected from the mathematical structure of the observable. Following Ref.~\cite{Neff2024}, the width dependence is fitted using the quadratic form of Eq.~(\ref{eq:bz}), with the resulting parameters presented in Fig.~\ref{fig:bz}. The amplitude $b$ remains negative and becomes more negative with decreasing RefMult, consistent with Fig.~\ref{fig:refmult}. The parameter $z$ exhibits no clear systematic dependence on centrality or beam energy, remaining near $140$--$143^\circ$ where well constrained, while its uncertainty increases substantially at 39~GeV due to the small amplitude.

The results in Figs.~\ref{fig:fixedN}--\ref{fig:bz} establish that the
UrQMD cascade produces a negative azimuthal-partition correlation with
a weak dependence on fixed proton multiplicity, a pronounced centrality
dependence, and a characteristic dependence on the partition width.
The following analysis examines which proton-pair populations carry
this correlation and where in the recorded reaction history its DD
component is already present.

\subsection{Final-state carriers of the negative correlation}

\begin{figure*}[t]
\centering
\includegraphics[width=\textwidth,keepaspectratio]
{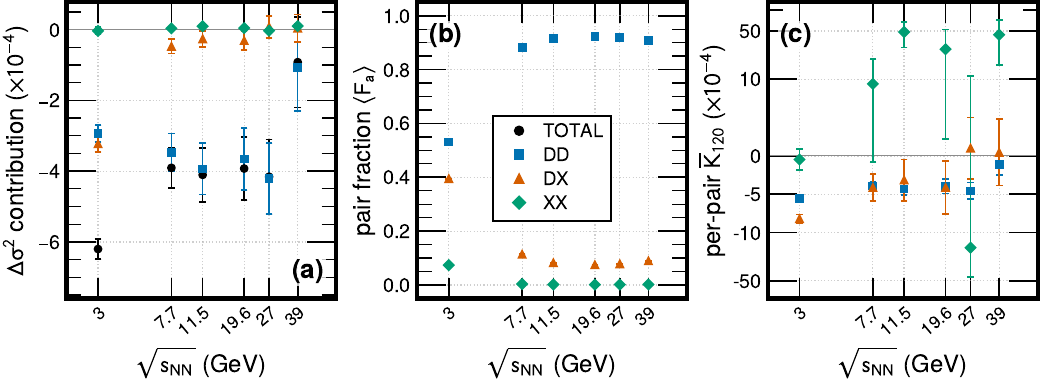}
\caption{\label{fig:carriers}
Final-state pair decomposition at $w=120^\circ$ for the 0--5\% proton-excluded RefMult class (80{,}000-event samples). (a)~Additive DD, DX, and XX contributions and their total. (b)~Mean pair fractions $\langle F_a\rangle$. (c)~Pair-weighted mean kernel $\overline{K}_{120}$ on a symmetric-log scale, linear within $\pm10\times10^{-4}$.}
\end{figure*}

Figure~\ref{fig:carriers}(a) shows the final-state pair decomposition
at $w=120^\circ$ for the 0--5\% proton-excluded RefMult selection of
the 80{,}000-event history-rich samples, which matches the centrality
of the benchmark results discussed above. The total pair contribution is resolved negative from 3 to 27~GeV and
consistent with zero at 39~GeV, mirroring the beam-energy dependence
of the benchmark result. The negative signal is carried by decay-involving pairs.
At 3~GeV the DD and DX contributions are both resolved and of
comparable magnitude, with DX slightly larger. From 7.7 to 27~GeV the
negative signal is dominated by the DD component, while DX and XX
remain small; a small positive XX contribution is resolved at
11.5~GeV, but its magnitude is an order of magnitude smaller than the
DD term and it does not carry the negative signal. At 39~GeV, the DD,
DX, and XX contributions, as well as their total, are all individually
consistent with zero. The weakening of the signal at this energy
therefore does not result from a cancellation between large components
of opposite sign.

Figure~\ref{fig:carriers}(b) and (c) separate the class abundance from
the mean pair-kernel correlation. From 7.7 to 27~GeV, DD pairs
constitute approximately 90\% of the accepted-proton pairs and exhibit
a negative per-pair kernel. However, these observations alone do not
establish a preferential negative association beyond the large DD pair
abundance. At 3~GeV, the
DX class combines a substantial pair fraction
($\langle F_{\rm DX}\rangle\simeq0.40$) with a more negative per-pair
kernel than DD, which explains why its additive contribution is
comparable to, and slightly larger than, the DD contribution. For the
XX class, the pair fraction stays below 0.5\% above 3~GeV, so its
per-pair kernel is only weakly constrained and its additive
contribution remains small. The dominance of decay-involving pairs
motivates tracing the DD population backward through the recorded
reaction history to determine where its azimuthal correlation is
already present.


\begin{figure*}[htbp]
\centering
\includegraphics[width=\textwidth,keepaspectratio]{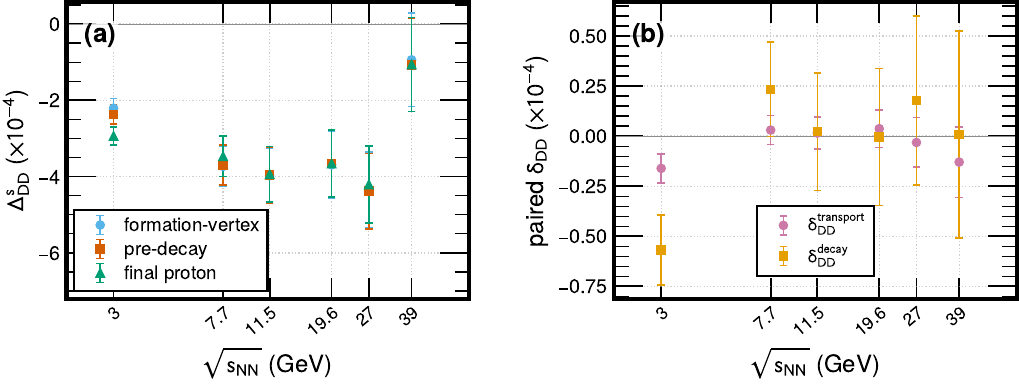}
\caption{\label{fig:stage}Reaction-history tracing of distinct-parent DD pairs versus $\sqrt{s_{NN}}$ for the 0--5\% proton-excluded RefMult class (80{,}000-event samples). (a)~Pair-kernel contributions $\Delta_{\rm DD}^{s}$ at $w=120^\circ$ for formation-vertex, pre-decay, and final-state momenta. (b)~Paired stage contrasts defined in Eq.~(\ref{eq:ddelta}).}
\end{figure*}

Figure~\ref{fig:stage}(a) follows the same distinct-parent DD pair
population through the three genealogical levels defined in
Sec.~\ref{sec:observable}, while Fig.~\ref{fig:stage}(b) shows the
corresponding paired differences of Eq.~(\ref{eq:ddelta}).
From 3 to 27~GeV, $\Delta_{\rm DD}$ is already resolved negative when
the parent momenta are evaluated at their respective formation
vertices and remains of similar magnitude before decay and in the
final-state protons. At 39~GeV, the contribution is consistent with
zero at all three levels.

From 7.7 to 39~GeV, both $\delta_{\rm DD}^{\rm transport}$ and
$\delta_{\rm DD}^{\rm decay}$ are statistically consistent with zero,
indicating no resolved modification of the DD pair-kernel correlation
between the formation-vertex and pre-decay levels or by the decay itself. At 3~GeV, both
contrasts are resolved negative,
$\delta_{\rm DD}^{\rm transport}=(-0.16\pm0.07)\times10^{-4}$ and
$\delta_{\rm DD}^{\rm decay}=(-0.57\pm0.18)\times10^{-4}$,
corresponding to small additional negative changes relative to the
already-negative formation-vertex contribution.

The reaction-history analysis therefore shows that for distinct-parent DD pairs, the negative pair-kernel correlation is already present in the parent momenta evaluated at their respective formation vertices over 3--27~GeV. No analogous formation-level statement is made for DX pairs, whose formation-level reconstruction is not supported by the persisted history records. At 39~GeV, both the formation-vertex contribution and the final-state carrier contributions are consistent with zero. These findings indicate that the final-state negative DD correlation is largely inherited from the angular correlations already present in the parent momenta at their formation vertices, providing an important constraint on its microscopic origin.

\begin{figure}[t]
\centering
\includegraphics[width=\columnwidth,keepaspectratio]
{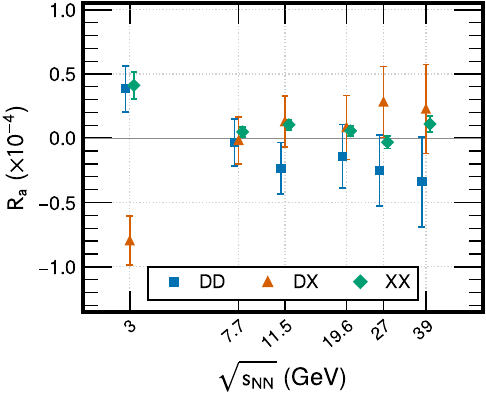}
\caption{\label{fig:nulltest} Residuals $R_a$ from the random-label null test for DD, DX, and XX pairs at $w=120^\circ$ in the 0--5\% proton-excluded RefMult class. The null test preserves the proton azimuthal configuration and the numbers of decay-fed and non-decay-fed protons in each event.}
\end{figure}

Figure~\ref{fig:nulltest} presents the residuals $R_a$ from the random-label null test defined in Sec.~\ref{sec:observable}. At 3~GeV, the DX residual is significantly negative, $R_{\rm DX}=(-0.80\pm0.19)\times10^{-4}$, while the DD and XX residuals are positive and statistically resolved. This indicates that the negative DX contribution exceeds the expectation based on pair abundance alone, whereas the DD and XX contributions are less negative than expected from their pair fractions. From 7.7 to 39~GeV, neither the DD nor the DX residuals differ significantly from zero, although the DD point estimates remain negative. Thus, the DD dominance at 7.7--27~GeV is consistent with its large pair fraction at the current statistical precision. At 11.5~GeV, the XX residual shows a small but significant positive deviation, $R_{\rm XX}=(+0.10\pm0.04)\times10^{-4}$. At 39~GeV, all three residuals are consistent with zero.

\section{Summary and Outlook}
\label{sec:summary}
In this work, we developed an LLM-assisted analysis framework that connects a general-purpose large language model to the UrQMD microscopic transport model through a model-specific UrQMD Skill and a deterministic scientific tool layer. The framework separates the LLM's role in scientific planning, workflow organization, and interpretation from numerical calculations, event-history reconstruction, and statistical analyses, which are performed by deterministic tools operating on recorded UrQMD output. This design enables the LLM to coordinate multi-step scientific analyses while ensuring that the underlying numerical procedures remain reproducible and traceable. We demonstrated the capabilities of this framework by investigating proton azimuthal-partition fluctuations in Au+Au collisions at $\sqrt{s_{NN}}=3.0$--39~GeV, combining final-state observable calculations with microscopic reaction-history tracing.

The UrQMD calculations exhibit negative azimuthal-partition correlations from 3 to 27~GeV in the 0--5\% centrality interval, while the signal is consistent with zero at 39~GeV. An exact pair decomposition reveals that the negative correlation is primarily associated with decay-fed protons, with DD pairs dominating at 7.7--27~GeV and both DD and DX contributions being important at 3~GeV. The random-label null test further identifies a nontrivial DX contribution at 3~GeV. Tracing the DD pairs through the recorded reaction history shows that their negative correlation is already present at the formation-vertex level from 3 to 27~GeV, with no statistically resolved modification during subsequent transport and decay at 7.7--39~GeV and small additional negative contributions at 3~GeV. These findings constrain the origin of the correlation without uniquely identifying its microscopic mechanism.

The results demonstrate the potential of the LLM-assisted framework for reproducible microscopic transport analyses and reaction-history tracing. Future work will investigate the contributions of different reaction sources, extend the framework to other transport models for systematic comparisons, and explore applications to experimental physics analyses, including low-energy heavy-ion experiments such as CEE~\cite{Lu:2016htm}. Incorporating detector acceptance and reconstruction effects would provide a pathway toward systematic comparisons between transport-model predictions and experimental observables.

\section*{Data Availability}
The numerical data supporting the results of this study, along with the source code and instructions for reproducing the main analyses, are publicly available in Ref.~\cite{LLM4UrQMD}.

\section*{AI Usage Statement}
The Kimi CLI agent orchestrated the UrQMD simulations and data generation through the UrQMD Skill and deterministic scientific tools. ChatGPT and DeepSeek were used to review and improve the readability and language of the manuscript. The authors independently verified all derivations, numerical results, and scientific statements and take full responsibility for the content.

\begin{acknowledgments}
We thank Prof. Long-Gang Pang of Central China Normal University for valuable suggestions and discussions. Most of this work was completed during a visit to Central China Normal University. F.-P. Li acknowledges support from the National Natural Science Foundation
of China (NSFC) under Grant Nos.~12325507, 12547102, and 12147101, and
from the National Key Research and Development Program of China under
Grant No.~2022YFA1604900. This work was also supported by the China
Postdoctoral Science Foundation under Grant No.~2025M783370.
\end{acknowledgments}

\bibliography{ref}
\end{document}